# Cavity enhanced UV combs generated by sum frequency mixing with near-IR chirped-pulse electro-optic combs for Rb atom sensing at 323 nm

**Jasper R. Stroud[1,*] and David F. Plusquellic[1,*]**

[1]*Applied Physics Division, Physical Measurement Laboratory, National Institute of Standards and Technology, Boulder, CO 80305*
**Corresponding author: jasper.stroud@nist.gov and david.plusquellic@nist.gov*

**A chirped-pulse electro-optic (EO) dual comb system operating near 821 nm is used to generate cavity enhanced dual combs in the ultraviolet (UV) region near 323 nm with an optical bandwidth of up to 90 GHz. The UV combs result from intracavity sum frequency mixing of the non-resonant near-IR combs with a cavity enhanced field at 532 nm in a nonlinear crystal. The cavity is pumped with 1 W at 532 nm and seeded with less than 10 mW in the near-IR region to generate a few µW of UV power within a detection bandwidth of < 20 MHz. The UV power is enhanced by 100-fold relative to single pass methods and is readily detectable using an avalanche photodiode. The system is used to measure the high-resolution UV comb spectra of Rb atoms near 323 nm (9 $^2P_{3/2}$←5 $^2S_{1/2}$). This method is easily extendable across the 300 nm to 400 nm region using telecom and near-IR dual EO comb sources for seeding.**

---

Mode-locked dual-comb spectroscopy (DCS) has proven to be an invaluable tool for the acquisition of high-resolution spectra over several THz of bandwidth [1,2,3]. Electro-optic dual comb sources (EO-DCS) generated from a single continuous wave laser source typically operate with reduced bandwidths of 100's of GHz but retain the high relative phase stability between combs necessary for phase coherent down conversion into the radio frequency (RF) domain [4,5,6,7]. Many atomic quantum sensing applications only require this smaller spectral coverage to enhance detection sensitivity. One challenge of CW electro-optic methods in the UV region is in the generation of the wavelength of interest. Direct modulation using electro-optic technology is robust for comb generation in the near-IR region from 700 nm to 2000 nm but becomes increasing difficult at shorter and longer wavelengths. The use of nonlinear frequency conversion methods based on difference frequency (DF) [8,9,10] or sum frequency (SF) [11,12,13,14] mixing is essential to transpose the EO-DCS for use in a wider range of applications. The extension into the UV region provides access to many atomic transitions of interest and unlocks new possibilities for quantum 2.0 techniques [15]. For example, highly coherent UV photon pump sources have been used for spontaneous parametric down conversion (SPDC) for generation of entangled pairs to probe atomic qubits [16]. Further, seed sources in the telecom region have been of particular interest to provide a low loss transmission method to interconnect qubit assemblies [17]. Hower, because of the requirement of single pass conversion of EO-DSCs, nonlinear methods often suffer from insufficient power in conversion for traditional detection.

Here we present a method to transpose dual electro-optic combs generated in the near-IR into the UV region with significantly enhanced conversion efficiency using a cavity-enhanced sum frequency (SF) mixing process. Figure 1 shows the simplified system diagram. About 1 W from a single frequency 532 nm Vanadate laser (Verdi-V18) is used to pump the resonant cavity while the remaining 10 W is used to pump a cavity stabilized titanium sapphire laser (Ti:Sapp, Coherent, 899-21, 720 nm to 980 nm [18]). The Ti:Sapp laser generates ≈300 mW near 815 nm of which 200 mW is coupled into a polarization preserving (PM) fiber. The near-IR seed light is split into two legs for input into fiber-coupled electro-optic phase modulators (EOM, EO-space, PM-5SE-10-PFA-PFA-850-LV, 3 dB insertion loss) in series with acousto-optic modulators (AOM, Brimrose, TEM-50-2-60-850-2FP, (5-6) dB insertion loss) with an overall 10 dBm insertion loss. The two legs are then recombined in fiber to form a dual comb interferometer. We note that the pump and Ti:Sapp lasers could easily be replaced with lower cost single frequency diode lasers.

The near-IR comb light from one leg (≈10 mW near 821 nm) is injected into a build-up cavity for non-resonant sum frequency mixing with the cavity enhanced 532 nm pump wave to produce coherent UV dual combs near 322.89 nm (vac). The dual comb UV output is passed through a 7 cm long cell to detect in absorption the $9^2P_{3/2}$ - $5^2S_{1/2}$ hyperfine transitions of $^{85}$Rb and $^{87}$Rb using a temperature-compensated silicon avalanche photodiode (Thorlabs, APD-130A2, 50 MHz, NEP≈0.75 pW/√Hz at 323 nm). The cavity-enhanced SF process therefore transposes to higher frequency (by 2.5-fold) the near-IR dual combs with sufficient power for photodiode detection, further extending the range and utility of EO-DCS systems.

As discussed in detail elsewhere [6], frequency chirped waveforms from an arbitrary waveform generator (AWG, Keysight M8195, 32 GS/s, 8-bit) are amplified to 1 W to drive the EOMs and define the comb tooth spacing and bandwidth of both the optical and down-converted RF waveforms. One of the waveforms includes a phase slip term to interleave the orders of the EOMs in the RF output. A small frequency difference of 4 kHz between the two AOMs serves to separate the positive and negative sidebands of the EOMs [6]. The two chirped waveforms are defined with the same chirp duration but with different bandwidths and mix to produce a RF chirped waveform at the difference in the chirp bandwidths (see supplement for details). For the first-order EOM sidebands, we down-convert ±15 GHz of optical bandwidth into 2 MHz of RF bandwidth for detection. For the interleaved higher orders of the EOM, the method enables the down-conversion of hundreds of GHz in the optical domain to a few tens of MHz in the radio frequency (RF) domain. In contrast to our previous results where frequency doubling near 840 nm was used to generate near-UV combs (≈ 420 nm) at twice the optical bandwidth [12], the SF process linearly transposes the near-IR bandwidth to the UV region. Further, unlike doubling (especially in the combined beam approach), the SF process does not generate unwanted SF mixing products that clutter the frequency doubled spectrum. We note that because of the phase matching limitations in BBO, wider comb bandwidths would be limited to < 200 GHz (i.e., 7$^{th}$ order) beyond which the SF power would diminish by 2-fold or more. Expansion beyond this limit would require, for example, cascaded EO modulator stages [19] together with nonlinear conversion in periodically poled lithium niobate as demonstrated for wavelengths > 330 nm [20,21].

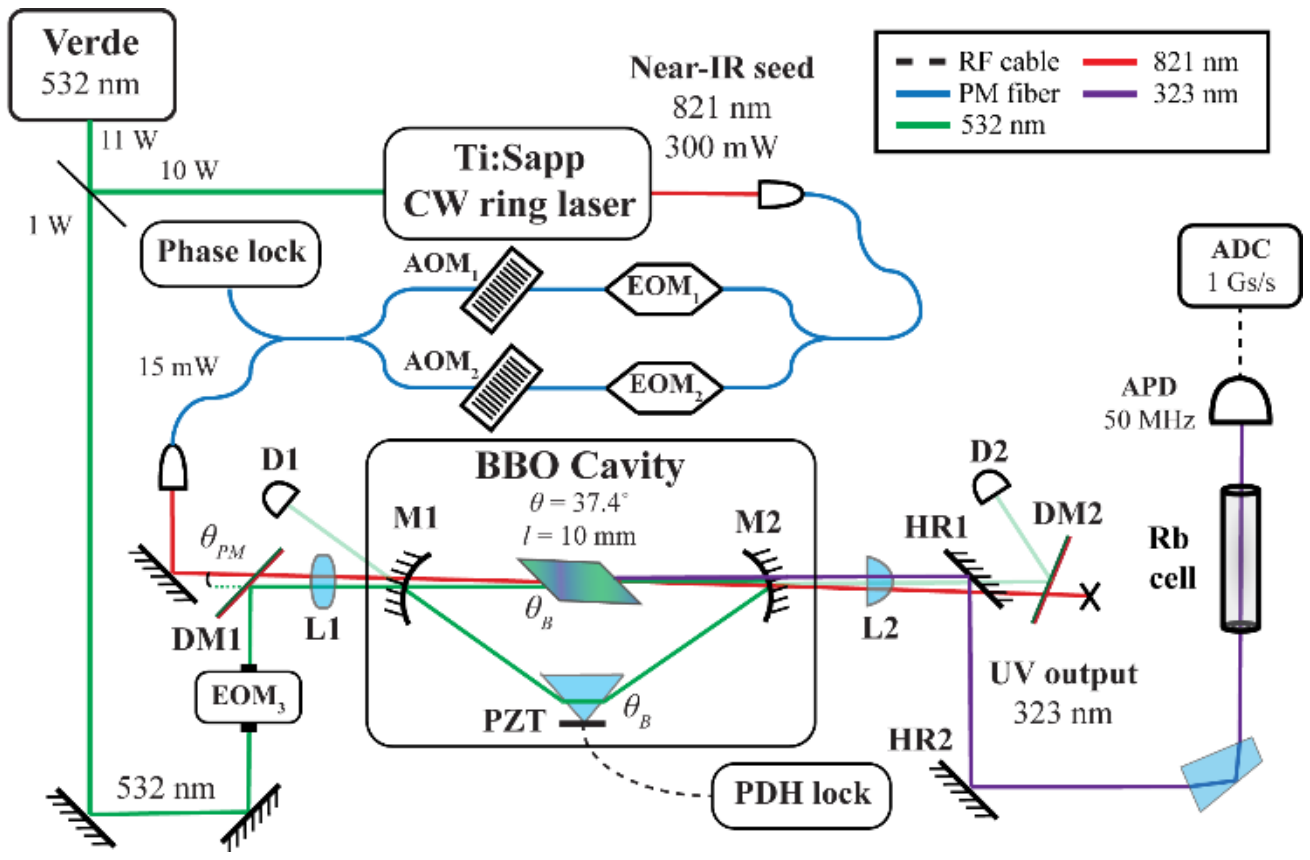

Fig. 1. The optical setup consisting of a 532 nm laser that is split to pump a Ti:Sapp laser used to generate near-IR dual combs (see text for details) for sum frequency mixing in BBO within a resonant cavity to generate UV combs. The 532 nm pump is mode matched into the buildup cavity using lens, L1, and resonance is maintained using a PDH lock for feedback to an intracavity prism. The UV combs are coupled out using mirror M2 and lens L2. A Pellin-Broca prisms removes residual pump and seed light prior to probing Rb atoms in absorption. D1: photodiode for PDH lock, D2: cavity fringe monitor, M1: input coupler for 532 nm, 99 % reflectivity, M2: 532 nm reflector, > 99.9 % and UV output coupler, L1: mode matching lens for 532 nm and 821 nm, L2: cylindrical lens for astigmatism compensation, DM1: long pass dichroic mirror, HR1 and HR2: dichroic mirrors for reflecting UV.

To enhance the SF conversion efficiency by orders of magnitude relative to our previous single pass methods [12], a commercial buildup cavity for frequency doubling (Spectra-Physic, Inc., Wavetrain II) is reconfigured for sum frequency generation [22,23,24]. Figure 1 shows the buildup cavity setup. About 1 W of 532 nm is mode matched into the cavity waste to enhance the intracavity power by ≈100-fold. The cavity has a free-spectral range of ≈1 GHz and a finesse near 100 that arises from the low round-trip loss of the pump wave from the Brewster angled surfaces (crystal + folding prism) within the cavity. The Vanadate pump laser is a single frequency source having a linewidth of < 10 MHz and is therefore well matched to the cavity line width for efficient mixing.

The 532 nm pump laser is mode matched to the cavity using a lens (L1) and is phase locked to the cavity using the Pound-Drever-Hall (PDH) method that includes a phase modulator and a photodiode (D1) to mix the reflected sideband modulation signal with the EO-driver reference for proportional-integral-differential (PID) feedback to a piezo transducer (PZT) attached to an intracavity prism for cavity length control. A second photodiode (D2) monitors the cavity transmission fringes to optimize pump alignment to the $TEM_{00}$ cavity mode. The dual comb seed light at ≈ 821 nm is combined with the pump using a long-pass dichroic mirror (DM1). Together with lens, L1, the near-IR light is mode matched to the cavity using an adjustable fiber collimator. The comb light is coupled through the cavity input coupler (M1, 99 % reflectivity at 532 nm) with ≈ 80 % efficiency and focused through the pump beam waste inside the BBO crystal (beta-barium borate, Conex, Brewster cut, 10 mm long, θ=37.3°, $\theta_B$=59.1°) for Type I non-colinear, non-critical phase matching. At a SF phase matching angle of 36.0°, the cavity outputs a few µW of UV comb light with > 50 % transmission efficiency through anti-Brewster crystal surface and the output coupler (M2, > 99.9% reflectivity at 532 nm) with orthogonal polarization to the pump and seed beams. The $d_{eff}$ of the mixing process is 1.95 pm/V with a UV beam walk-off of 78 mrad. The cylindrical lens (L2) largely removes the astigmatism of the UV beam from the Brewster cut crystal. The 323 nm dual comb light is separated from the pump and seed beams using two dichroic mirrors (HR) and a Pellin Broca prism. The UV dual comb power of a few µW is attenuated by 3-fold to prevent saturation of the avalanche photodiode ($P_{sat}$≈1.5 µW).

Figure 2a shows the time domain interferogram of the dual comb system. The 1 ms total record consists of twenty 50 µs chirped pulses that span the down-converted RF region from 3 MHz to 5 MHz in 1$^{st}$ order. The overall modulation arises from the 4 kHz beat note between the two AOMs. The insets in Fig. 2a show the 50 µs chirp and the transition region between the chirped pulses. The AWG waveform amplitudes are set to enhance the 2$^{nd}$ and 3$^{rd}$ order sidebands of the EOMs, which span 60 GHz and 90 GHz in the near-IR (and UV), respectively. Figure 2b shows the 1$^{st}$, 2$^{nd}$, and 3$^{rd}$ order combs in the frequency domain. The RF comb tooth spacing is 20 kHz (the inverse of the 50 µs chirp repetition period) and shown for both sidebands in the inset. The positive and negative sidebands are separated by the 4 kHz beat note, while the orders are separated by a linearly increasing phase shift applied to the 20 repeated chirps.

The dual UV comb output is coupled through a temperature controlled 75 mm long Rb reference cell that is heated to approximately 165 °C using heaters around the cell (the windows are kept ≈ 10 °C higher). At this vapor pressure (≈ 1.3 Pa), the fractional absorption of the strongest Rb line is near 20 %. The light is then focused on a 1 mm in diameter APD using a short focal length lens. The APD output is digitized at 1 GS/s by an analog to digital converter (ADC).

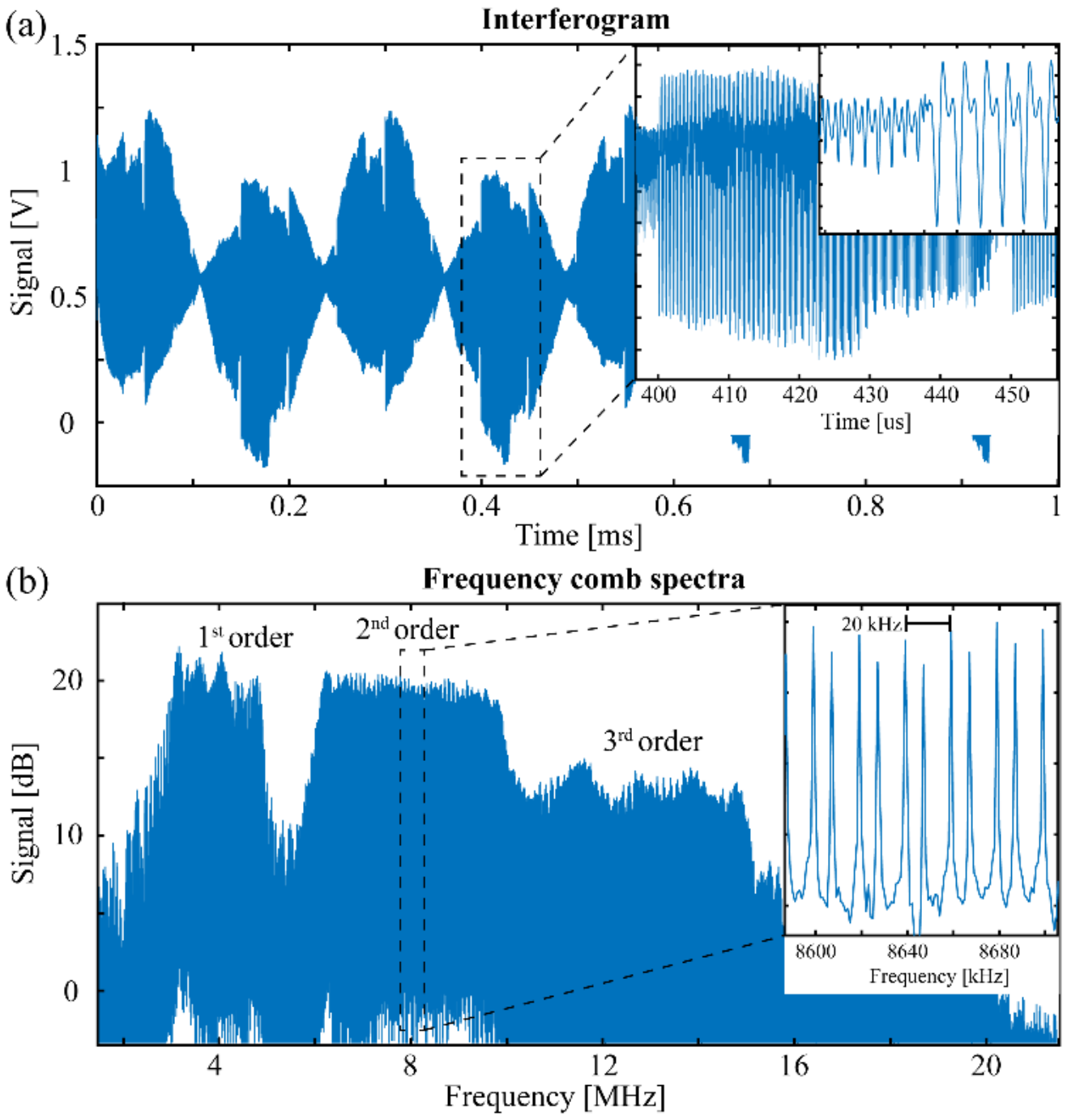

Fig. 2. (a) The time domain interferogram showing the 4 kHz beat note with the insets showing expanded portions of the down-converted 50 µs chirped pulse. (b) The frequency domain spectrum consists of the 1st order RF comb spanning from 3 MHz to 5 MHz together with the higher harmonics. The expanded inset shows the 2nd-order comb tooth separation.

Unlike weak-signal detection schemes relying on photomultiplier tubes [11,12], the strong analog output signal from the APD is also used to phase lock the 4 kHz beat note between the two legs of the near-IR interferometer through feedback to one of the AOM drives [6].

Improvements in the software and hardware have enabled both the time domain interferogram and the comb magnitude and phase spectra to be simultaneously saved in real time. The multi-threaded program consists of 44 threads that each collect 1 ms of data with 80 % throughput. The 44 buffers are coherently averaged in the time domain and Fourier transformed (44 ms interval). Following 100 averages (N), the coherent time domain (4.4 s interval) and averaged frequency domain representations of the signal are saved. These two methods are found to produce nearly identical spectra which verifies the robust phase lock of the interferometer and suggests √N averaging over the ≈ 5 s timescale. Further √N improvements in the SNR ratio were not found after ≈ 5 s as detailed in the supplemental. However, a dual beam or a quantum correlated measurement approach [25] utilizing the 821 nm comb (not performed here) will likely lead to further improvements. We also note that while a photon counting sampling rate of 1 GS/s was used here, the RF detection bandwidth needed (< 50 MHz) at these high light levels can be acquired at 1/10 of this sampling rate.

The data are processed to generate both time domain and frequency domain spectra using background measurements acquired in an adjacent off-resonance region. The 2nd order spectra are shown in Fig. 3a and 3b where the time domain data show the $^{85}$Rb/$^{87}$Rb spectrum around 322.9 nm having four Doppler broadened spectral features, each consisting of several unresolved hyperfine components. To remove the effects of temporal magnification [6], the frequency domain comb spectra are first back-transformed and fit in the time domain [26]. Figures 3c and 3d show that the 2nd and 3rd order spectra, respectively, are nearly identical in the time domain except for a slight increase in the 3rd order rms noise in the residuals (x10). However, the frequency domain spectra show increased residuals (x2) that contain additional ripples from temporal magnification, the source of which is currently under investigation.

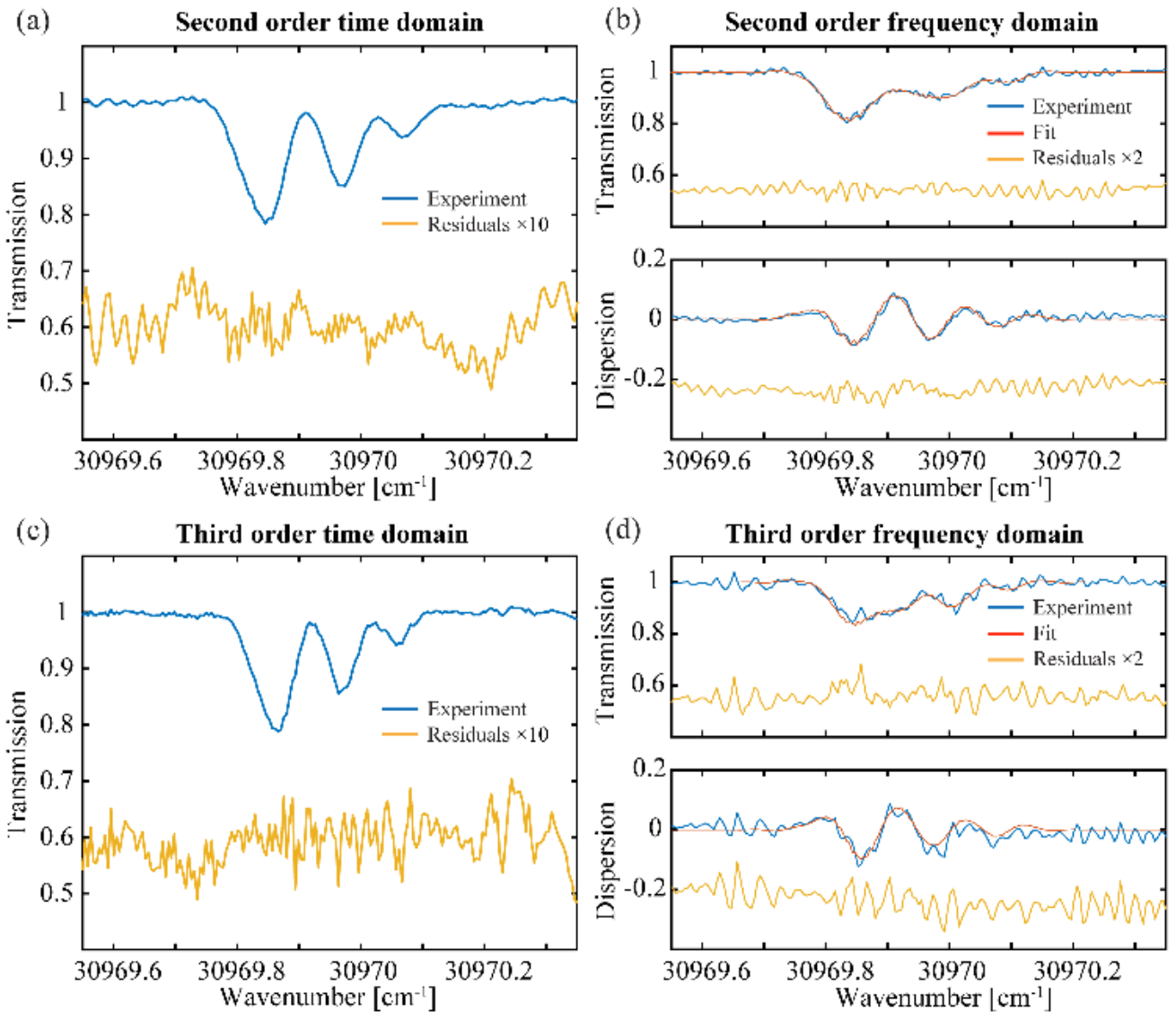

Fig. 3. The time domain transmission spectra of Rb (9 $^2P_{3/2}$←5 $^2S_{1/2}$) obtained in (a) 2nd order and (c) 3rd order with scaled residuals shown in yellow. The corresponding frequency domain spectra are shown in (b) and (d) where relative to the model in red, show increased ripples in the residuals. The phase spectra only appear in the frequency domain because of the differential chirped pulse scheme (see ref. [12]).

The best fit line center frequencies and line strengths are shown in Table 1 using three independent data sets obtained for both the 2nd and 3rd order spectra. A slowly varying background was first removed, and the fits were performed by floating the four Doppler broadened ($\Delta\nu_{FWHM}$ ≈ 1.5 GHz) line centers and line strengths calculated from the estimated vapor pressure (1.3 Pa) [27] in the 7.5 cm long Rb reference cell. The unresolved hyperfine components in each of the four lines were not included in the model. The absolute line positions have estimated uncertainties of ±2 GHz at 532 nm + 821 nm from wavemeter calibration measurements using a stabilized HeNe laser. Their average position is in good agreement with previous work [28]. The relative frequency uncertainties range from 12 MHz to 35 MHz, where the largest error occurs for the unresolved $^{87}$Rb (F"=2) line appearing as a shoulder on the $^{85}$Rb (F"=3) feature.

The long-term stability for continuous comb acquisition depends on the quality of the RF beat note lock and the drift rate of the 532 nm and 821 nm lasers. The beat note lock is robust for extended periods in tests of up to 6 hrs although as noted above, phase errors after 5 s are not corrected by this system. The slow hourly drift of the pump laser at 532 nm is < ± 300 MHz. A high precision wavemeter (Toptica, High Finesse, WS7) is used to monitor both the 532 nm and 821 nm laser frequencies for SF calibration. The relative uncertainties in Table I reflect the estimated ± 20 MHz precision of this method for all independent measurements performed over a 1 hr period. Active feedback corrections to the Ti:Sapp laser frequency could further improve the relative precision to < ± 3 MHz [29].

Table 1. Best-fit relative line center frequencies and line strengths of the 9 $^2P_{3/2}$←5 $^2S_{1/2}$ transitions of the $^{87}$Rb and $^{85}$Rb isotopologues. Uncertainties shown in the least significant digits are determined from the RMS differences (Type B, k=1 or 1σ) in six independent spectral fits of the four main-line features. The absolute frequency uncertainties are estimated at ±2 GHz (Type B, k=1 or 1σ) (see text for details).

| Ground state | Frequency (THz) | Line Strength (cm$^{-1}$/mol cm$^{-2}$ ×10$^{-17}$) |
|---|---|---|
| $^{87}$Rb F" = 2 | 928. 458677(35) | 1.12(4) |
| $^{85}$Rb F" = 3 | 928. 459799(12) | 2.52(9) |
| $^{85}$Rb F" = 2 | 928. 462816(22) | 1.85(11) |
| $^{87}$Rb F" = 1 | 928. 465429(23) | 0.67(9) |

The line strength uncertainties are < 6 % except for the weakest line. The significantly higher temperature used here (165 °C) vs that used in our previous study of Rb at 420 nm (113 °C) [12] offsets for the ≈ 27-fold difference in line strengths measured in these two regions.

In this work, dual electro-optical frequency combs generated in the near-IR (≈ 821 nm) are linearly transposed into the UV region (≈ 323 nm) by sum frequency generation with the cavity enhanced field at 532 nm in a nonlinear crystal, thereby enhancing the UV conversion efficiency by ≈ 2 orders of magnitude relative to single pass methods. The UV power of a few µWs enables the detection of the down-converted RF comb with a low cost APD detector. While a wavelength near 323 nm was chosen for Rb atom sensing, the UV tuning range of the system can easily be extended to cover the 300 nm to 400 nm region using EO-combs generated over the region from 700 nm to 1600 nm. We note that nearly identical results were obtained for Rb in recent measurements across the 300 nm to 400 nm region using difference frequency generation of entangled pairs at 266 nm and will be reported elsewhere.

**Disclosures.** The authors declare no conflicts of interest.

**Data availability.** Data will be made available.